# Atomic Structures of Riboflavin (Vitamin B2) and its Reduced Form with Bond Lengths Based on Additivity of Atomic Radii

Raji Heyrovska

Institute of Biophysics, Academy of Sciences of the Czech Republic.

Email: rheyrovs@hotmail.com

## Abstract

It has been shown recently that chemical bond lengths, in general, like those in the components of nucleic acids, caffeine related compounds, all essential amino acids, methane, benzene, graphene and fullerene are sums of the radii of adjacent atoms constituting the bond. Earlier, the crystal ionic distances in all alkali halides and lengths of many partially ionic bonds were also accounted for by the additivity of ionic as well as covalent radii. Here, the atomic structures of riboflavin and its reduced form are presented based on the additivity of the same set of atomic radii as for other biological molecules.

## Introduction

An introduction to the additivity of radii in bond lengths can be found in [1-6], where [1a] concerns the interionic distances in all alkali halides and many other bond lengths, [1b] shows the results for nearly forty metal hydrides. The additivity of radii is shown [2] to hold even for hydrogen bonds in inorganic and biochemical groups. The chapter in [3] summarizes these and includes the results for aqueous solutions. In [4], the skeletal bond lengths in the molecular components of nucleic acids are shown to be the sums of the appropriate radii of carbon, nitrogen, oxygen, phosphorus and hydrogen atoms. Likewise in [5] and [6] are presented the results for caffeine related molecules and essential amino acids (which also contain sulfur) respectively. The latest work [7] is on the comparison of the atomic structures of methane, benzene, graphene [7a] and fullerene [7b]. This work presents for the first time, the structures of riboflavin and its reduced form also based on the same set of atomic radii as for all the other biological molecules mentioned above.

## Atomic structures of riboflavin (vitamin B2) and its reduced form

The conventional molecular structure [8] of riboflavin, also known as vitamin B2, is shown in Fig. 1, and an introduction to the properties of this molecule, can be found in [9]. All the bonds shown in Fig. 1 are sums of the appropriate radii of the adjacent atoms, which are given in Fig. 2 (these radii are the same as those in [2-5]). The structure of riboflavin at the atomic level can be seen in Fig. 2, where all the atoms have their specific contribution to the relevant bond lengths. Similarly, the atomic structure of the reduced form of riboflavin is shown in Fig. 3, where all interatomic distances are sums of the concerned atomic radii.

**Acknowledgements:** The author thanks the IBP for financial support and Professor E. Palecek of the IBP, Academy of Sciences of the Czech Republic for his moral support.

**Fig. 1.** The conventional molecular structure of riboflavin [8].

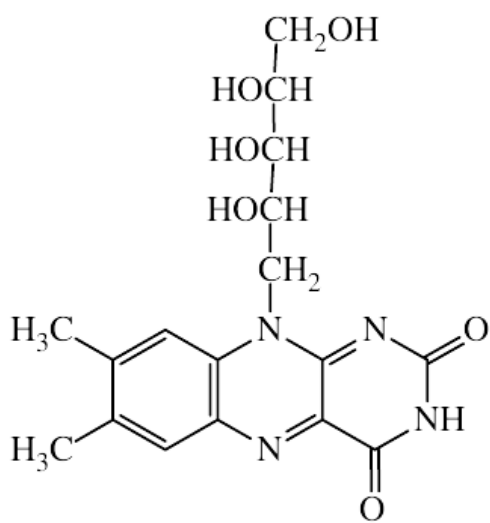

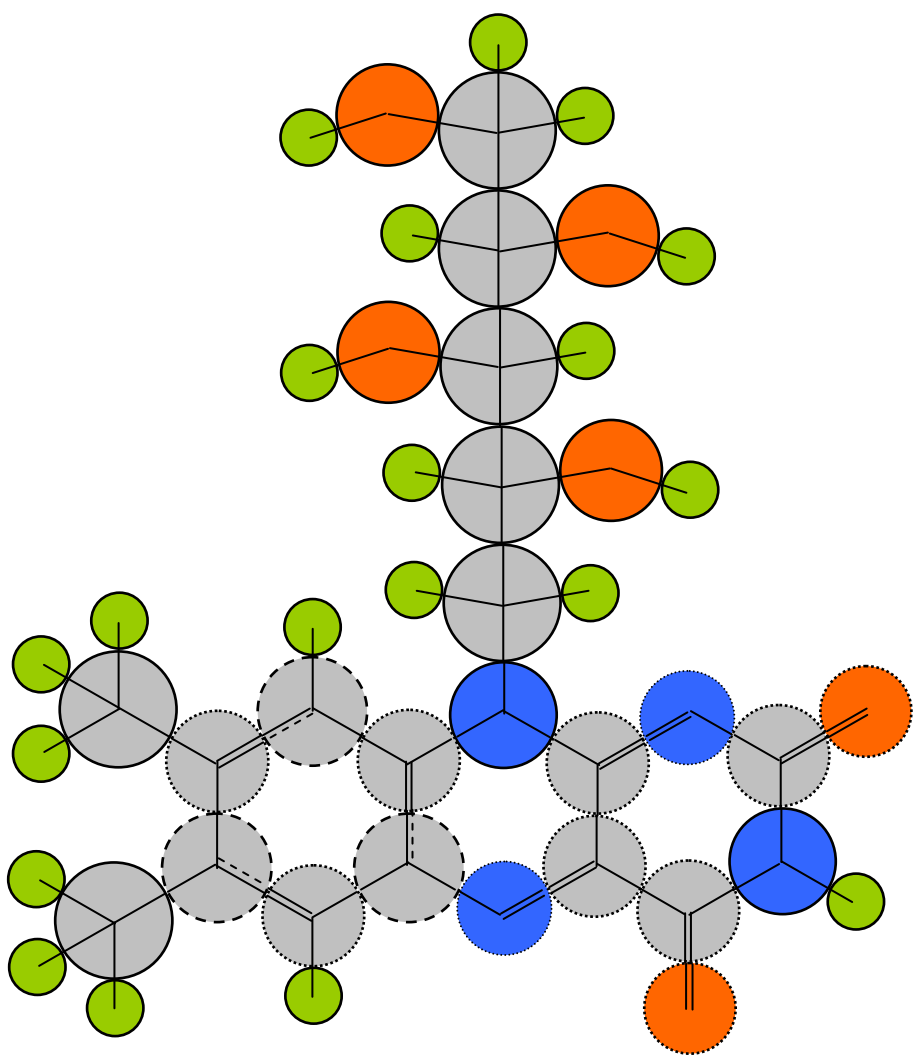

**Fig. 2. Atomic structure of riboflavin.** Radii (R = half the interatomic distance between atoms of the same kind) of carbon (C), nitrogen (N), oxygen (O) and hydrogen (H) atoms. (Subscripts: s.b.: single bond, d.b.: double bond and res.: resonance bond).

**Fig. 3. Atomic structure of the reduced form of riboflavin.** Radii (R = half the interatomic distance between atoms of the same kind) of carbon (C), nitrogen (N), oxygen (O) and hydrogen (H) atoms. (Subscripts: s.b.: single bond, d.b.: double bond and res.: resonance bond).

| $C_{s.b.}$ | $C_{res.}$ | $C_{d.b.}$ | $N_{s.b.}$ | $N_{d.b.}$ | $O_{s.b.}$ | $O_{d.b.}$ | $H_{s.b.}$ |
|---|---|---|---|---|---|---|---|
| R: 0.77 | 0.72 | 0.67 | 0.70 | 0.62 | 0.67 | 0.60 | 0.37 Å |

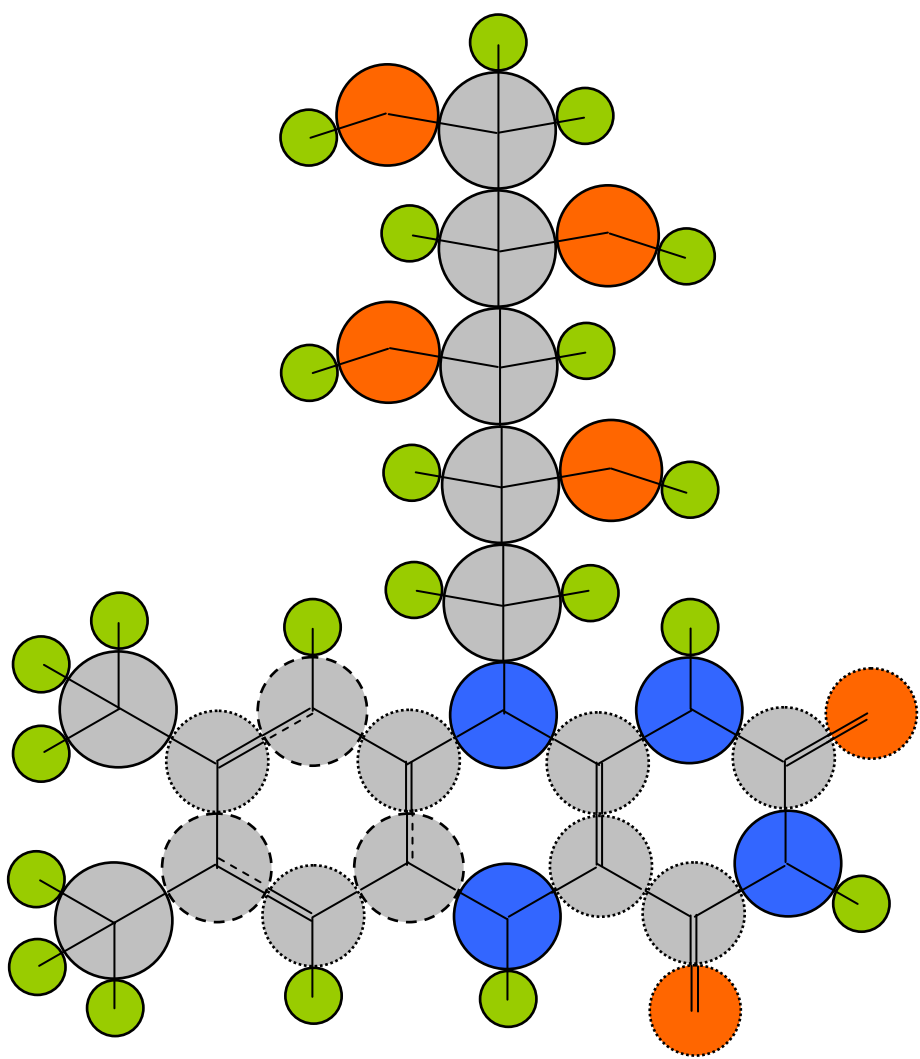